\documentclass[9pt,twocolumn]{IEEEtran}

\usepackage[english]{babel}
\usepackage{epsfig}
\usepackage{dsfont}
\usepackage{mathrsfs}
\usepackage[mathscr]{eucal}
\usepackage{graphicx}
\usepackage{amsmath,amsfonts,amsthm,amssymb,amsbsy,amsopn,amstext}
\usepackage{latexsym}
\usepackage{color,multicol,makeidx}
\usepackage{CJK}
\usepackage{indentfirst}
\usepackage{bm}
\usepackage{flushend}
\usepackage{balance}
\usepackage{mathrsfs}
\usepackage{float}
\usepackage{ccaption}
\usepackage[caption=false,font=normalsize,labelfont=sf,textfont=sf]{subfig}
\usepackage{multirow}

\newtheorem{them}{Theorem}
\newtheorem{defn}{Definition}
\newtheorem{lem}{Lemma}
\newtheorem{rem}{Remark}

\ifCLASSINFOpdf
\else
\fi
\begin{document}
%
\title{Distributed Surrounding Design of Target Region
with Complex Adjacency Matrices}

\author{Youcheng~Lou
        and~Yiguang~Hong
\thanks{This work is
supported by the NNSF of China under Grant 61174071.}         
\thanks{The authors are with
Academy of Mathematics and Systems Science, Chinese Academy of Sciences, Beijing 100190, China.
       Email: {\tt\small  louyoucheng@amss.ac.cn, yghong@iss.ac.cn}}
}

%

\maketitle

\emph{Abstract}--This is a complete version of the 6-page IEEE TAC technical note \cite{lou15}.
In this paper, we consider the distributed surrounding of a convex target
set by a group of agents with switching communication
graphs. We propose a distributed controller to surround a given
set with the same distance and desired projection
angles specified by a complex-value adjacency matrix.
Under mild connectivity assumptions, we give results in both consistent and inconsistent cases for
the set surrounding in a plane.  Also, we provide sufficient conditions for the
multi-agent coordination when the
convex set contains only the origin.
\vskip 2mm

%
\IEEEpeerreviewmaketitle

\section{Introduction}
The distributed coordination and control of multi-agent systems has
been investigated from various perspectives due to its various
applications.
After the study of consensus or formation of multi-agent
systems \cite{olf, renbook, hu10, chen11,
anderson, lin}, much attention has been paid to set
coordination problems of multi-agent systems.
Among the studies of multi-agent set coordination,
distributed containment control has achieved much, which makes
agents reach a convex set maybe spanned by multiple leaders
\cite{shi09, caoren, louhong, shi12}.  Moreover, some results were obtained to control a group of
agents in order to protect or surround a convex target set. For
example, the distributed controller was designed for the agents to
surround all stationary leaders in the convex hull spanned by the
agents in \cite{chenf}, while a model was provided for
multiple robots to protect a target region \cite{earl}.  However,
many theoretical problems to surround a target set remain to be
solved.

On the other hand, complex Laplacians or rotation matrices have been
applied to consensus and formation (see \cite{lin1,lin2,ren}), partially because the complex representation may significantly simplify the analysis when the state space is a plane.
Formation control for directed acyclic graphs with complex Laplacians
and related stability analysis were discussed in \cite{lin2}, while
new methods were developed for pattern formation with complex-value
elements in \cite{lin1}.

The objective of this paper is to study the distributed set
surrounding design based on complex adjacency matrices, that is, to design a
distributed protocol to make a group of agents protect/surround a
convex set in a plane. We first propose a distributed controller to
make all agents achieve the set projection with the same distance
and different projection angles specified by a given complex-value adjacency
matrix. For uniformly jointly strongly connected undirected graph
and fixed strongly connected graph, we provide the initial conditions
guaranteeing that all agents will not converge to the set. Then we
investigate the special case when the set becomes the origin, with
a necessary and sufficient condition in the fixed
strongly connected graph case.
In addition, our results also extend some existing ones including the
consensus \cite{olf, renbook} and bipartite consensus \cite{alt}.

The contributions of this paper include: 1) we proposed a
distributed controller to solve the set surrounding problem under
the switching communication graphs; 2) we characterize the
relationship between the consistency of directed cycles of the
configuration graph and the system dynamic behavior, or roughly
speaking, the consistent cycles produce the consistent case, while
inconsistent cycles yield the inconsistent case; 3)
we extend some existing results of consensus and bipartite consensus when the set contains only one point.

The paper is organized as follows. Section II gives preliminary
knowledge and the problem formulation. Section III provides the main
results for the distributed set surrounding problems and then considers
an important special case when the target becomes the origin.
Then Section IV gives a numerical
example for illustration.  Finally, Section V shows some concluding remarks.

\emph{Notation}: $\mathbb{R}$ and $\mathbb{C}$ denote the real
field and complex field, respectively; $|\cdot|$ denotes the modulus
of a complex number or the number of elements in a set; $P_{X}(\cdot)$ denotes the
projection operator onto the closed convex set $X$; $z^p$ denotes the projection vector of point $z$ onto $X$, i.e., $z^p=z-P_{X}(z)$;
$|\cdot|_{X}$ denotes the distance
 between a point and $X$, i.e., $|z|_{X}=|z-P_{X}(z)|$; $\iota=\sqrt{-1}$ denotes the imaginary unit;
 $\angle z$ denotes the argument of complex number $z$;
$\langle \cdot,\cdot\rangle$ denotes the
inner product of two complex numbers, i.e., $\langle a_1+a_2\iota, b_1+b_2\iota\rangle=a_1b_1+a_2b_2$.

\section{Preliminaries and Formulation}

In this section, we first introduce preliminary knowledge and then
formulate the distributed set surrounding problem.

\subsection{Preliminaries}

A digraph (or directed graph)
$\mathcal{G}=(\mathcal{V},\mathcal{E})$ consists of node set
$\mathcal{V}=\{1,2,...,n\}$ and arc set $\mathcal{E}\subseteq
\mathcal{V} \times \mathcal{V}$ \cite{God}. A weak path in digraph $\mathcal{G}$ is an alternating sequence
$i_1e_1i_2e_2\cdots i_{k}e_{k}i_{k+1}$ of nodes
$i_r,r=1,...,k+1$ and arcs $e_r=(i_r, i_{r+1})\in \mathcal{E}$ or
$e_r=(i_{r+1}, i_r)\in \mathcal{E},r=1,...,k;$ if
$e_r=(i_r,i_{r+1})$ for all $r$, the weak path becomes a
directed path; if $i_1=i_{k+1}$, the weak path is called a weak
cycle; A weak cycle containing a directed path is called a directed
cycle. Digraph $\mathcal{G}$ is said to be weakly strongly connected if there
exists a weak path in $\mathcal{G}$ between every pair of nodes
in $\mathcal{V}$, and strongly connected if there
exists a directed path in $\mathcal{G}$ between every pair of nodes
in $\mathcal{V}$. Moreover, $\mathcal{G}$ is undirected if $(i,j)\in \mathcal{E}$ is equivalent to
$(j,i)\in \mathcal{E}$.
Undirected graph (digraph) $\mathcal{G}$ is said to be a
(directed) tree if there is one node such that there is one and only
one (directed) path from any other node to this node. Undirected graph (digraph) $\mathcal{G}$ is said to contain
a (directed) spanning tree if it has a (directed) tree containing all nodes of $\mathcal{G}$ as
its subgraph.
 Here we assume $\mathcal{G}$ contains no self-loop, i.e.,
$(i,i)\not\in \mathcal{E}$, $i\in\mathcal{V}$.

Consider a multi-agent system consists of $n$ agents.
Let
$\sigma:[0,\infty)\rightarrow\mathcal{Q}$ be a piecewise constant
function to describe the switching graph process with $\mathcal{Q}$
the index set of all possible digraphs on $\mathcal{V}$.
Denote a switching graph with signal
$\sigma$ as
${\mathcal{G}}_{\sigma}=(\mathcal{V},\mathcal{E}_{\sigma})$, which is called a \emph{communication} graph to describe the (time-varying) communication between the agents (regarded as nodes) with taking its connection weight $a_{ij}=1$ if $(i,j)\in \mathcal{E}_\sigma$ for simplicity.
Denote $\mathcal{G}_{\sigma}([s_1,s_2))$ as the union graph with node set
$\mathcal{V}$ and arc set $\bigcup_{s_1\leq t<
s_2}\mathcal{E}_{\sigma(t)}$, $0\leq s_1<s_2$.
Let $\{t_k,k\geq0\}:=\Delta$ with $t_0=0$ denote the set of all switching moments of switching graph $\mathcal{G}_{\sigma}$. The switching
digraph $\mathcal{G}_{\sigma}$ is uniformly jointly strongly
connected $(UJSC)$ if there exists $T>0$ such that
$\mathcal{G}_{\sigma}([t,t+T))$ is strongly connected for any $t\geq
0$. As usual, we assume there is a dwell time as the lower bound
between two consecutive switching moments. That is, there exists $\tau>0$ such that
$t_{k+1}-t_k\geq \tau$ for all $k$.

The Dini derivative of a continuous function $f:
(a, b)\rightarrow \mathbb{R}$ at $t\in(a, b)$ is defined as
follows:
$$
D^+f(t)=\limsup_{s\rightarrow 0^+}\frac{f(t+s)-f(t)}{s}.
$$
Clearly, $f$ is non-increasing on $(a,b)$ if $D^+f(t)\leq0$, $\forall t\in(a,b)$. The
following result can be found in \cite{dan}.

\begin{lem}\label{lem2}
Let $f_i(t,x): \mathbb{R}\times \mathbb{R}^m\rightarrow \mathbb{R}$, $i=1,...,M$ be continuously differentiable
and $f(t,x)=\max_{1\leq i\leq M}f_i(t,x)$. Then $D^+f(t,x(t))=\max_{i\in
\mathcal{I}(t)}\dot f_i(t,x(t))$, where
$\mathcal{I}(t)=\big\{i|f_i(t,x(t))=f(t,x(t)),1\leq i\leq M\big\}$.
\end{lem}

As we know, a set $K$ is said to be convex if
$(1-\lambda)z_1+\lambda z_2\in K$ whenever $z_1, z_2\in K$ and
$0\leq\lambda \leq1$.  Moreover, let $P_K(\cdot):
\mathbb{C}\rightarrow K$ be the projection operator onto closed
convex set $K$, i.e., $P_K(z)$ is the unique element in $K$
satisfying $\inf_{y\in K}|z-y|=|z-P_K(z)|:=|z|_{K}$ \cite{rock}.
In the following convergence analysis, we need to extend the Barbalat¡¯s Lemma to
a switching case.

\begin{lem}\label{mod}
Let $g:[0,\infty)\rightarrow \mathbb{R}$ be a continuous function
with $\lim_{t\rightarrow\infty}g(t)=g_0$.
Suppose $g$ is continuously differential except all the switching moments $t_k,t\geq0$
and the derivative $\dot g$ is uniformly continuous with respect to all time intervals $(t_k,t_{k+1})$ (i.e.,  for any $\varepsilon>0$, there exists $\delta>0$ such that
for any $k$ and any $s_1,s_2$ satisfying $t_k<s_1<s_2<t_{k+1}$,
when $|s_2-s_1|\leq\delta$, $|\dot g(s_2)-\dot g(s_1)|\leq\varepsilon$).
Then $\lim_{t\rightarrow\infty}\dot g(t)=0$.
\end{lem}

\emph{Proof:} This conclusion
can be shown by contradiction via the almost same arguments in the proof of the well-known
Barbalat's Lemma. Without loss of generality, suppose $\limsup_{t\rightarrow\infty}\dot g(t):=g^*>0$.
Then let $\{s_k\}_{k\geq0}$ be a subsequence such that  $g^*/2\leq\dot g(s_k)\leq 3g^*/2$ for all $k$.
Therefore, for $\varepsilon=g^*/4$, there exists $\delta>0$ such that for any $k$ and any $r_1,r_2$ satisfying $t_k<r_1<r_2<t_{k+1}$,
when $|r_2-r_1|\leq\delta$, $|\dot g(r_2)-\dot g(r_1)|\leq\varepsilon$.
Since $t_{k+1}-t_k\geq\tau>0$ for all $k$, we can assume $\delta$ is sufficiently small such that
for each $k$, either $s_k+\delta\in(t_k,t_{k+1})$ or $s_k-\delta\in(t_k,t_{k+1})$.
First suppose $s_k+\delta\in(t_k,t_{k+1})$. Then
$\dot g(s_k+\delta)\geq\dot g(s_k)-g^*/4\geq g^*/4$. As a result,
$g(s_k+\delta)=g(s_k)+\int^{s_k+\delta}_{s_k}\dot g(t)dt\geq g(s_k)+\delta g^*/4$.
Similarly, we also have $g(s_k-\delta)\geq g(s_k)+\delta g^*/4$  when $s_k-\delta\in(t_k,t_{k+1})$.
This contradicts the hypothesis $\lim_{t\rightarrow\infty}g(t)=g_0$.
Thus, the conclusion follows. \hfill$\square$

\subsection{Problem Formulation}

Consider the $n$ agents described by the first-order integrator
\begin{equation}
\label{1}
 \dot x_i(t)=u_i(t),\quad i=1,...,n,
\end{equation}
where $x_i, u_i\in \mathbb{C}$ are the state and control input of
agent $i$ in the plane, respectively.  Consider a 2-dimensional
bounded closed convex set $X\subseteq \mathbb{R}^2$ to be surrounded. For a
desired surrounding configuration or pattern, we need to assign the desired
relative projection angles between the agents when they surround
$X$.  To this end, we give a complex-value adjacency matrix $W=(w_{ij})\in
\mathbb{C}^{n\times n}$ to describe the desired relative angles of
projections for agents to $X$ as follows: $w_{ii}=0$, $i=1,...,n$
and either $|w_{ij}|=1$ or $ w_{ij}=0$ for $i\neq j$. In this
way, we get a digraph $\mathcal{G}^w=(\mathcal{V},\mathcal{E}^w)$
with $\mathcal{E}^w=\{(i,j)|w_{ij}\neq0\}$, which is called a {\em
configuration} graph. Meanwhile, $w_{ij}$ is called the
\emph{configuration weight} of arc $(i,j)\in\mathcal{E}^w$.
A weak cycle $i_1e_1i_2e_2\cdots i_{k}e_{k}i_1$ in
$\mathcal{G}^w$ is said to be consistent if
\begin{align}
 \prod^{k}_{r=1}w(e_r)=1,\nonumber
\end{align}
where $w(e_r)= w_{i_ri_{r+1}}$ for $e_r=(i_r, i_{r+1})$,
$w(e_r)= w^{-1}_{i_{r+1}i_r}$ for $e_r=(i_{r+1}, i_r)$;
otherwise, it
is said to be inconsistent.  Clearly, $w_{ij}w_{ji}=1$ in the consistent case when
$w_{ij}\neq 0$ and $w_{ji}\neq 0$.

\begin{rem}
Although no convex set gets involved in the control design in
multi-agent formation \cite{anderson, lin1}, its design is directly
based on the desired formation configuration determined by the
desired relative distances or positions.  Sometimes, the desired
formation can be described by a set of desired relative position
vectors $d_{ij}$ to show the desired position of agent $j$ relative
to that of agent $i$ for $i,j=1,...,n$. In this case, for a given weak
cycle $i_1e_1i_2e_2\cdots i_{k}e_{k}i_1$, we also have the
consistent case with $\sum_{r=1}^k d(e_r)=0$, and inconsistent case
with $\sum_{r=1}^k d(e_r)\neq0$, where $i_{k+1}=i_1$, $d(e_r)=
d_{i_ri_{r+1}}$ for $e_r=(i_r, i_{r+1})$, $d(e_r)= -d_{i_{r+1}i_r}$
for $e_r=(i_{r+1}, i_r)$. It is known that the formation may fail in
the inconsistent case.  In our problem, the desired relative
projection angles are described by $w_{ij}$ to achieve the desired
surrounding configuration, which plays a similar role as $d_{ij}$ in
the formation.  Therefore, in both formation and surrounding
problems, the agents' indexes are given in the desired
configuration.
\end{rem}

In this paper, we consider how to surround the given set $X$ with
the same distance by the $n$ agents from different projection angles
(that is, the rotation angles of projection vectors) specified by $W$.
To be strict, we introduce the following definition.

\begin{defn}
\label{def1}  The distributed set surrounding is achieved for system
(\ref{1}) with a distributed control $u_i$ if, for any
initial condition $x_i(0), 1\leq i\leq n$,
$$
\lim_{t\to\infty}w_{ij}(x_j(t)-P_{X}(x_j(t)))-(x_i(t)-P_{X}(x_i(t)))=0
$$
for $(i,j)\in \mathcal{E}^w$.
\end{defn}

In fact, there are two cases for the set surrounding:
\begin{itemize}
\item {\bf Consistent case}:
All agents surround the convex set $X$ with the same nonzero distance to $X$
and desired projection angles between each other determined by the
entries $w_{ij}$ of $W$.
\item {\bf Inconsistent case}: all agents converge to the convex set $X$.
\end{itemize}

In what follows, we will show: if the weights given in the configuration graph are inconsistent, the
inconsistent case appears; if the weights are consistent, we can achieve the consistent case somehow.

\begin{rem}
Different from the surrounding formulation given in \cite{chenf},
the agents in our problem not only surround the target set but also
keep the same distance from the target set (potentially for balance
or coordination concerns). Additionally, the inconsistent case
resulting from the inconsistency of configuration graph is related to
containment problems \cite{shi12, louhong, caoren}.
\end{rem}

In practice, node $i$ may not receive the information from node $j$
sometimes due to
communication failure or energy saving. Denote the set of all arcs
transiting information successfully at time $t$ as
${\mathcal{E}}_{\sigma(t)}$, which is a subset of $\mathcal{E}^w$,
and the resulting graph is
${\mathcal{G}}_{\sigma(t)}=(\mathcal{V},{\mathcal{E}}_{\sigma(t)})$, which is the
\emph{communication} graph of the multi-agent system.
Note that the
configuration graph $\mathcal{G}^w$ shows the desired relative projection angles of
the agents, while the communication graph ${\mathcal{G}}_{\sigma}$, a subgraph of $\mathcal{G}^w$, describes the
communication topology of the agents.
 Let
$\mathcal{N}_i=\{j|(i,j)\in\mathcal{E}^w\}$ and $\mathcal{N}_i(\sigma(t))\subseteq \mathcal{N}_i$ denote the
neighbor set of node $i$ in communication graph ${\mathcal{G}}_{\sigma(t)}$.
Then we take the following control
\begin{align}
\label{2}
u_i(t)&=\sum_{j\in \mathcal{N}_i(\sigma(t))}\big[w_{ij}\big(x_j(t)-P_{X}(x_j(t))\big)\nonumber\\
&\qquad\qquad\qquad-\big(x_i(t)-P_{X}(x_i(t))\big)\big].
\end{align}
As usual, in the design of controller (\ref{2}), agents only count
in the received information from their neighbors.

\begin{rem}
Let us check the role of complex-value adjacency matrix $W$. In Definition \ref{def1}, the complex-value configuration weight $w_{ij}(=e^{\alpha_{ij}\iota})$ indicates that $\alpha_{ij}$ is the desired angle difference between
projection vector of agent $i$ onto set $X$
and that of agent $j$.
Because $|w_{ij}|=1$ for $(i,j)\in \mathcal{E}^w$,
$w_{ij}x^p_j(t)-x^p_i(t)\rightarrow 0$ implies
$|x_i(t)|_{X}-|x_j(t)|_{X}\rightarrow 0$, and therefore, all agents will have the same distance to
the convex set
when the (consistent) set surrounding is achieved.  If $|w_{ij}|\neq 1$, we may get the set surrounding with different distances from the agents to the target set.
\end{rem}

\section{Main Results}

In this section, we will solve the following basic surrounding problems: (i)
How to design distributed controllers to achieve the set
surrounding? (ii) What initial conditions can guarantee the consistent
case? (iii) What happens when the target set consists of only one point?

Before we study the set surrounding problem, we first show that the consistent case of the set surrounding problem is well-defined, which can be achieved in some situations.

\begin{them}
\label{them0} Consider a complex-value adjacency matrix $W$ and the resulting configuration graph $\mathcal{G}^w=(\mathcal{V},\mathcal{E}^w)$.
If all weak cycles of $\mathcal{G}^w$
are consistent, then there are $z_i,i=1,...,n$ such that $|z_i|_{X}\neq0,\;\forall i$
and $z_i-P_{X}(z_i)=w_{ij}(z_j-P_{X}(z_j)),\;\forall (i,j)\in\mathcal{E}^w$.
\end{them}

\emph{Proof:}
For any $(i_1,i_0)\in\mathcal{E}^w$ and $z_{i_0}\not\in X$, take
$e_{i_1i_0}=w_{i_1i_0}(z_{i_0}-P_{X}(z_{i_0}))$. Define a
hyperplane
$$
\mathcal{H}_{\lambda}=\big\{z|\;\langle z, e_{i_1i_0}\rangle=
\langle P_{X}(z_{i_0})+\lambda e_{i_1i_0},
e_{i_1i_0}\rangle\big\},\;\lambda\geq0.
$$
and denote the two corresponding closed half spaces as
$\mathcal{H}^+_{\lambda}$ and $\mathcal{H}^-_{\lambda}$,
respectively. Let
$$
\lambda^*=\sup\big\{\lambda|\lambda\geq0,\;X\bigcap\mathcal{H}^+_{\lambda}\neq \emptyset\big\}.
$$
Since $P_{X}(z_{i_0})\in\mathcal{H}^+_{0}$ and
$X\bigcap\mathcal{H}^+_{\lambda}=\emptyset$ for sufficiently
large $\lambda$, $\lambda^*<\infty$. It is easy to see that
$X\bigcap\mathcal{H}^+_{\lambda^*}\neq\emptyset$ and
$X\subseteq\mathcal{H}^-_{\lambda^*}$. Let
$z_{i_1}=y_{i_1}+e_{i_1i_0}$ with $y_{i_1}\in
X\bigcap\mathcal{H}^+_{\lambda^*}$. Clearly, $z_{i_1}\not\in X$,
$P_{X}(z_{i_1})=y_{i_1}$ and
$z_{i_1}-P_{X}(z_{i_1})=e_{i_1i_0}=w_{i_1i_0}(z_{i_0}-P_{X}(z_{i_0}))$.

If $\mathcal{G}^w$ contains no weak cycle, we can apply the
similar arguments to all the other arcs in $\mathcal{E}^w$ to obtain the
conclusion; if $\mathcal{G}^w$ contains weak cycles and all its weak cycles are consistent, we can continue the above procedures until there
are $z_1,....,z_n$ such that
$z_i-P_{X}(z_i)=w_{ij}(z_j-P_{X}(z_j))$ for all
$(i,j)\in\mathcal{E}^w_*$, where $(\mathcal{V},\mathcal{E}^w_*)$ is a
maximal spanning subgraph of $\mathcal{G}^w$ containing no weak
cycle.  Because all weak cycles of $\mathcal{G}^w$ are consistent,
$z_i-P_{X}(z_i)=w_{ij}(z_j-P_{X}(z_j))$ also holds for all the
other $w_{ij}$ when $(i,j)\in\mathcal{E}^w\backslash\mathcal{E}^w_*$.
Thus, the proof is completed. \hfill$\square$

In the following two subsections, we will show that inconsistent cycles
yield the inconsistent case for any initial conditions and the consistent cycles imply the consistent case for
all initial conditions except a bounded set, respectively.   Then in the third subsection, we will reveal the inherent relationships between consensus and our problem with the set containing only one point.

\subsection{Set surrounding}

The following results provide sufficient conditions for the considered set surrounding problem.

\begin{them}
\label{them1} (i) The distributed set surrounding is achieved for system
(\ref{1}) with control law (\ref{2}) if the communication graph
$\mathcal{G}_{\sigma}$ is $UJSC$ and all directed cycles of the
configuration graph $\mathcal{G}^w$ are consistent;

(ii) $\lim_{t\to\infty}|x_i(t)|_{X}=0$, $i=1,...,n$ for any initial conditions if
the communication graph $\mathcal{G}_{\sigma(t)}\equiv\mathcal{G}^w$ is fixed, strongly
connected and there are inconsistent weak cycles in $\mathcal{G}^w$.
\end{them}

\emph{Proof:}
(i) Define the arc set connecting infinitely long time
$$
\mathcal{E}_\propto=\big\{(i,j)|\exists\;\{s_k\}_{k\geq0}, s_k\rightarrow\infty\;\mbox{such that}\; (i,j)\in\mathcal{E}_{\sigma(s_k)} \big\}
$$
and corresponding graph $\mathcal{G}_\propto=(\mathcal{V},\mathcal{E}_\propto)$.
Clearly, $\mathcal{G}_\propto$ is a subgraph of the configuration graph $\mathcal{G}^w$.

Define
$$
d(t)=\max_{1\leq i\leq
n}d_i(t),\;\;d_i(t)=\frac{1}{2}|x_i(t)|^2_{X},\;i\in
\mathcal{V},t\geq 0,
$$
which are nonnegative. According to Proposition 1 in \cite{aub} (page 24),
$|x_i(t)|^2_{X}$ is continuously differentiable and its derivative
is $2\langle x^p_i(t), \dot x_i(t)\rangle.$   Applying Lemma
\ref{lem2} gives
\begin{align}\label{13}
D^+d(t)&=\max_{i\in \mathcal{I}(t)}\big\langle x^p_i(t),\dot x_i(t)\big\rangle\nonumber\\
&=\max_{i\in \mathcal{I}(t)}\Big\langle x^p_i(t),\sum_{j\in \mathcal{N}_i(\sigma(t))}\big(w_{ij}x^p_j(t)-
x^p_i(t)\big)\Big\rangle\nonumber\\
&\leq\max_{i\in \mathcal{I}(t)}\sum_{j\in \mathcal{N}_i(\sigma(t))}\big(|x_i(t)|_{X}|x_j(t)|_{X}-
|x_i(t)|^2_{X}\big)\nonumber\\
&\leq0
\end{align}
with $\mathcal{I}(t)=\big\{j|j\in\mathcal{V},d_j(t)=d(t)\big\}$ and $t\not\in \Delta$.
Therefore, it follows from (\ref{13}) that $d(t)$ is non-increasing and then converges to a finite
number, that is,
\begin{align}\label{limit}
\lim_{t\rightarrow\infty}d(t)=d^*.
\end{align}
As a result, the agent states
$x_i(t),i\in\mathcal{V},t\geq0$ are
bounded because $X$ is bounded.
In addition, if $d^*=0$, the conclusion is obvious. Suppose $d^*>0$ in the following proof
of this part.

Since the switching communication graph $\mathcal{G}_{\sigma}$ is $UJSC$, by
similar procedures in the proof of Lemma 4.3 in \cite{shi2}, we can
show that $\lim_{t\to\infty}d_i(t)=d^*$, $i=1,...,n$.  Based on the
boundedness of system states and the non-expansive property of projection operator,
we can find that $\dot
d_i$ is uniformly continuous with respect to time intervals $[t_k,t_{k+1})$ for $k\geq0$. Then by Lemma
\ref{mod} we have $\lim_{t\rightarrow\infty}\dot
d_i(t)=0$, that is,
\begin{align}
\lim_{t\rightarrow\infty}\dot d_i(t)&=
\lim_{t\rightarrow\infty}\Big\langle x^p_i(t),\sum^n_{j=1}\chi_{ij}(t)\big(w_{ij}x^p_j(t)-
x^p_i(t)\big)\Big\rangle\nonumber\\
&=\lim_{t\rightarrow\infty}\sum^n_{j=1}
\chi_{ij}(t)\Big(-|x_i(t)|^2_{X}\nonumber\\
&\quad\quad+|x_j(t)|_{X}|x_i(t)|_{X}
\cos(\angle w_{ij}x^p_j(t) -\angle x^p_i(t))\Big)\nonumber\\
&=\lim_{t\rightarrow\infty}\sum^n_{j=1}\chi_{ij}(t)\Big(-2d^*\nonumber\\
&\qquad\qquad\qquad\qquad+2d^*
\cos(\angle w_{ij}x^p_j(t) -\angle x^p_i(t))\Big)\nonumber\\
&=0,\nonumber
\end{align}
which implies
\begin{align}\label{6}
\lim_{t\rightarrow\infty,\;t\in\Xi_{ij}}w_{ij}x^p_j(t)-x^p_i(t)
=0,
\end{align}
where $\Xi_{i,j}=\{t|(i,j)\in\mathcal{E}_{\sigma(t)}\}$,
$$
\chi_{ij}(t)=\left\{
\begin{array}{ll}
1,\;\mbox{if}\;\;(i,j)\in\mathcal{E}_{\sigma(t)};\\
0,\;\; \mbox{otherwise}.\\
\end{array}
\right.
$$
It follows from
(\ref{6}) that for any $\varepsilon>0$, there is $T_1>0$ such that, when
$t\geq T_1$, $|w_{ij}x^p_j(t)-x^p_i(t)|\leq \varepsilon$ for
$(i,j)\in \mathcal{E}_{\sigma(t)}$, and then $|\dot
x_i(t)|\leq(n-1)\varepsilon$. Thus,
$|x_i(t_2)-x_i(t_1)|\leq\int^{t_2}_{t_1}|\dot x_i(s)|ds
\leq(n-1)(t_2-t_1)\varepsilon$, and then
$|x^p_i(t_2)-x^p_i(t_1)|\leq 2(n-1)(t_2-t_1)\varepsilon$
for $t_2\geq t_1\geq T_1$, where the last inequality follows from
the non-expansive property of projection operator:
$|P_{X}(z_1)-P_{X}(z_2)|\leq |z_1-z_2|,\forall z_1,z_2$.
Without loss of generality, we assume $T_1$ is a sufficiently large number such that
$\mathcal{E}_{\sigma(t)}\subseteq\mathcal{E}_\propto,$ $\forall t\geq T_1$.

Take $(i_0,j_0)\in \mathcal{E}_\propto$ arbitrarily. Since the
union graph $\mathcal{G}_{\sigma}([t,t+T))$ is strongly connected,
there exist nodes $i_1,...,i_k, k\leq n-2$ and time instants $t\leq
s_0,s_1,...,s_k<t+T$ such that
$(i_r,i_{r+1})\in\mathcal{E}_{\sigma(s_r)},r=0,...,k-1$ and
$(i_k, j_0)\in\mathcal{E}_{\sigma(s_k)}$. At the same time, there
also exists a directed path $\mathcal{P}$ from $j_0$ to $i_0$ in
$\mathcal{G}_{\sigma}([t,t+T))$. Denote the product of all configuration weights
on $\mathcal{P}$ as $w_*$.

Since $\mathcal{G}_{\sigma}([t,t+T))$ is a subgraph of $\mathcal{G}^w$ and
all directed cycles of $\mathcal{G}^w$ are consistent, all directed cycles of
$\mathcal{G}_{\sigma}([t,t+T))$ are also consistent. Therefore,
$\prod^{k-1}_{r=0}w_{i_ri_{r+1}}w_{i_kj_0}w_*=1$. Moreover, since
$i_0(i_0,j_0)j_0\mathcal{P}$ is a directed cycle in $\mathcal{G}^w$, $w_{i_0j_0}w_*=1$.
Thus, $\prod^{k-1}_{r=0}w_{i_ri_{r+1}}w_{i_kj_0}=w_{i_0j_0}$ and
\begin{align}
&\ \ \ |x^p_{i_0}(t)-w_{i_0j_0}x^p_{j_0}(t)|\nonumber\\
&=\big|x^p_{i_0}(t)-
\prod^{k-1}_{r=0}w_{i_ri_{r+1}}w_{i_kj_0}x^p_{j_0}(t)\big|\nonumber\\
&\leq|x^p_{i_0}(t)-x^p_{i_0}(s_0)|+
|x^p_{i_0}(s_0)-w_{i_0i_1}x^p_{i_1}(s_0)|\nonumber\\
&\quad+|w_{i_0i_1}x^p_{i_1}(s_0)-w_{i_0i_1}x^p_{i_1}(s_1)|
\;+\cdots\nonumber\\
&\quad+\Big|w_{k-2}x^p_{i_{k-1}}(s_{k-1})-
w_{k-1}x^p_{i_k}(s_{k-1})\Big|\nonumber\\
&\quad+\Big|w_{k-1}x^p_{i_k}(s_{k-1})-
w_{k-1}x^p_{i_k}(s_{k})\Big|\nonumber\\
&\quad+\Big|w_{k-1}x^p_{i_k}(s_{k})-
w_{i_kj_0}w_{k-1}x^p_{j_0}(s_k)\Big|\nonumber\\
&\quad+\Big|w_{i_kj_0}w_{k-1}x^p_{j_0}(s_k)-
w_{i_kj_0}w_{k-1}x^p_{j_0}(t)\Big|\nonumber\\
&\leq (k+1)\varepsilon+2(k+2)(n-1)^2T\varepsilon\nonumber\\
&\leq (n-1)\varepsilon+2n(n-1)^2T\varepsilon,\nonumber
\end{align}
where $w_{q}=\prod^{q}_{r=0}w_{i_ri_{r+1}}$. Since $\varepsilon$ can be
sufficiently small, we can further obtain
\begin{align}\label{ineq}
\lim_{t\to\infty} w_{ij}x^p_j(t)-x^p_i(t)=0,\;\;\forall(i,j)\in\mathcal{E}_\propto.
\end{align}
Clearly, due to the uniformly strong connectivity of $\mathcal{G}_{\sigma}$,
$\mathcal{G}_{\propto}$ is strongly connected. Combining the previous conclusion, (\ref{ineq})
and the consistency of directed cycles of $\mathcal{G}^w$, we have
$\lim_{t\to\infty} w_{ij}x^p_j(t)-x^p_i(t)=0$ for all the other $w_{ij}$ with
$(i,j)\in\mathcal{E}^w\backslash\mathcal{E}_\propto$.
Thus, the proof of part (i) is completed.

(ii) We first show this conclusion for the case when there exist
inconsistent directed cycles in $\mathcal{G}^w$. Let $i_1e_1i_2e_2\cdots
i_ke_ki_1$ be an inconsistent directed cycle in $\mathcal{G}^w$ with
\begin{align}\label{12}
\prod^{k}_{r=1}w_{i_ri_{r+1}}\neq1,
\end{align}
$i_{k+1}=i_1$. From (\ref{6}), we have
$\lim_{t\rightarrow\infty}w_{i_ri_{r+1}}x^p_{i_{r+1}}(t)-x^p_{i_r}(t)=0$ for $1\leq r\leq k$.
Therefore,
$$
\lim_{t\rightarrow\infty}x^p_{i_1}(t)\Big(1-\prod^{k}_{r=1}w_{i_ri_{r+1}}\Big)=0,
$$
which implies $\lim_{t\rightarrow\infty}d_{i_1}(t)=0$ and then
$d^*=0$. For the case of existing weak cycles (not directed
cycles) in $\mathcal{G}^w$, we can similarly show this conclusion by
replacing the configuration weight $w_{i_ri_{r+1}}$ in (\ref{12}) with
$w^{-1}_{i_{r+1}i_r}$ corresponding to arc $e_r=(i_{r+1}, i_r)$.

Thus, we complete the proof. \hfill$\square$

From the proof of Theorem \ref{them1}, we can find that the
conclusion (ii) also holds under the following relaxed connectivity
condition: the communication graph $\mathcal{G}_{\sigma}$
is $UJSC$ and there exist a time sequence $\{s_k\}^\infty_{k=0},s_k\rightarrow\infty$ and
$b_0>0$ such that $\mathcal{E}^w\subseteq
\bigcup^{s_k+b_0}_{t=s_k}\mathcal{E}_{\sigma(t)}$ for $k\geq0$.

\begin{rem}
In the consistent set surrounding, the distance of all agents surrounding the convex set $X$
is same. In fact, we can consider a more general set surrounding problem that all agents surround the convex set X with
the pre-specified relative distance relationship to $X$ among agents determined by some matrix $(d_{ij})$
($d_{ij}$, $(i,j)\in\mathcal{E}$ describes the proportion of the desired distance of agent $i$ onto set $X$
and that of agent $j$ onto set $X$) and desired projection angles
between each other determined by the entries $w_{ij}$ of $W$. We can find that this new set
surrounding problem can be solved by distributed controller (\ref{2}) by replacing $w_{ij}$
with $d_{ij}w_{ij}$ if the communication graph
$\mathcal{G}_{\sigma}$ is $UJSC$, and the consistence definition of directed cycles of the
configuration graph $\mathcal{G}^w$ with replacing $w_{ij}$
with $d_{ij}w_{ij}$ holds.
\end{rem}

%

\begin{rem}
Here we provide another simple proof of Theorem \ref{them1} (i)
when $w_{ij}=1$ for all $i,j$ and the communication graph is a fixed undirected connected graph.
First of all, by $\nabla|x|^2_{X}=2(x-P_X(x))$, we have
$$
\frac{d|x_i(t)|^2_{X}}{dt}=2\big\langle x^p_i(t), \sum_{j\in \mathcal{N}_i}\big(x^p_j(t)-
x^p_i(t)\big)\big\rangle.
$$
Taking the sum for the two sides of the above equality over $i=1,...,n$, by the undirectedness of the graph $(\mathcal{V},\mathcal{E})$,
we have
\begin{align}
\label{add}\frac{d\sum^n_{i=1}|x_i(t)|^2_{X}}{dt}&=-2\sum_{(i,j)\in \mathcal{E}}|x^p_j(t)-x^p_i(t)|^2\\
&\leq0,\nonumber
\end{align}
which implies that $\sum^n_{i=1}|x_i(t)|^2_{X}$ is non-increasing and then its limit exists.
As a result, the system states $x_i(t),i,t\geq0$ are bounded.
This combines with (\ref{add}) lead to
$$
\int^\infty_{0}\sum_{(i,j)\in \mathcal{E}}|x^p_j(t)-x^p_i(t)|^2dt<\infty.
$$

From the boundedness of system states, we can see that $\sum_{(i,j)\in \mathcal{E}}|x^p_j(t)-x^p_i(t)|^2$
is uniformly continuous on $[0,\infty)$. Then it follows from the Babalat's Lemma that
$\lim_{t\rightarrow\infty}\sum_{(i,j)\in \mathcal{E}}|x^p_j(t)-x^p_i(t)|^2=0$.
This implies that $\lim_{t\rightarrow\infty}(x^p_j(t)-x^p_i(t))=0$ for all $(i,j)\in \mathcal{E}$
and then the set surrounding is achieved.
\end{rem}

\subsection{Consistent Case}

Theorem \ref{them1} showed that
the distributed set surrounding can be achieved under \emph{UJSC}
communication graph condition.  In this subsection,
we further show under the case without inconsistent cycles of
$\mathcal{G}^w$, how to select initial conditions such that the
consistent case (that is, $d^*>0$ given in (\ref{limit})) can be
guaranteed.

Let $L_{\sigma(t)}$ be the matrix with entries
\begin{equation}
\label{lap}
(L_{\sigma(t)})_{ij}=\left\{
  \begin{array}{ll}
     |\mathcal{N}_i(\sigma(t))|, & \mbox{if}\;\;i=j; \\
    -w_{ij}, & \mbox{if}\;\;i\neq j,\;j\in \mathcal{N}_i(\sigma(t));\\
     0,   &\mbox{otherwise}.
  \end{array}
\right.
 \end{equation}
Then system (\ref{1}) with control law (\ref{2}) can be written in the following compact form
\begin{equation}
\label{9}
\dot x(t)=-L_{\sigma(t)}x^p(t),
 \end{equation}
where $x(t)=(x_1(t),...,x_n(t))^T$ is the stack vector
of agents' states, $x^p(t)=(x^p_1(t),...,x^p_n(t))^T$ is the stack vector
of agents' projection vectors.

We first consider system (\ref{9}) with a $UJSC$ undirected graph
$\mathcal{G}_{\sigma}$, where all directed cycles of the configuration graph $\mathcal{G}^w$
are consistent. Without loss of generality (otherwise we can relabel
the index of nodes), we take a spanning tree $\mathcal{T}$ of $\mathcal{G}^w$ as follows:
$\mathcal{T}=\bigcup^\rho_{k=1}\mathcal{T}_k$, where the initial and
the terminal nodes of the path $\mathcal{T}_k$ are $i_k$ and 1,
respectively; the nodes in the path from $i_k$ to 1 are in the order
$i_k,i_k-1,\ldots, i_{k-1}+1, 1,$ $1\leq k\leq \rho$, $i_0=1$.
Associated with the $n$ nodes, we define $n$ nonzero complex numbers
\begin{align}
p_{1}=1,\;p_{j}=w^{-1}_{(i_{k-1}+1)1}\prod^{j-1}_{r=i_{k-1}+1}w^{-1}_{(r+1)r}\;
\mbox{for}\;i_{k-1}+1\leq j\leq i_k.\nonumber
\end{align}
Denote a diagonal matrix
\begin{equation}
\label{pp} P={\rm diag}(p_{1},...,p_{n})
\end{equation}
with diagonal elements $p_i,1\leq i\leq n$.
It is easy to see that
$\widetilde{L}_{\sigma(t)}=PL_{\sigma(t)}P^{-1}$ is the Laplacian\footnote{
The Laplacian $\bar L$ of a digraph $\mathcal{G}=(\mathcal{V},\mathcal{E})$ is defined as:
$(\bar L)_{ii}=|\bar{\mathcal{N}}_i|$, $(\bar L)_{ij}=-1$ for $j\neq i,j\in\bar{\mathcal{N}}_i$
and all other entries are zero, where $\bar{\mathcal{N}}_i=\{j|(i,j)\in\mathcal{E}\}$ \cite{God}.}
 of
undirected graph $\mathcal{G}_{\sigma(t)}$. Then we have

\begin{them}
\label{them5} In the switching $UJSC$ undirected graph case, $d^*>0$ if
the initial condition $x(0)$ satisfies
$\big|\textbf{1}^TPx(0)/n\big|>\sup_{z\in X}|z|$.
\end{them}

\emph{Proof:} Let $\widetilde{x}(t)=Px(t)$. Clearly, system
(\ref{9}) can be written as
\begin{equation}
\dot {\widetilde{x}}(t)=-\widetilde{L}_{\sigma(t)}Px^p(t).\nonumber
 \end{equation}
Because $\textbf{1}^T\widetilde{L}_{\sigma(t)}\equiv0$ with
$\textbf{1}=(1,...,1)^T$, $ \textbf{1}^T\widetilde{x}(t)/n$ is
time-invariant. Note that $\sup_{z\in X}|z|$ is a finite number since
$X$ is bounded.

We prove the conclusion by contradiction. Hence suppose $d^*=0$. Since
$\lim_{t\rightarrow\infty}d_i(t)=d^*$ under
the $UJSC$ assumption, $\lim_{t\rightarrow\infty}|x_i(t)|_{X}=0$.
Therefore, $\limsup_{t\to\infty}|x_i(t)|\leq \sup_{z\in X}|z|$ and
then $\big|\textbf{1}^T\widetilde{x}(t)/n\big|\leq \sup_{z\in
X}|z|$, which yields a contradiction due to
$\textbf{1}^T\widetilde{x}(t)/n\equiv\textbf{1}^T\widetilde{x}(0)/n$.
\hfill$\square$

Next we consider system (\ref{9}) under a fixed strongly connected
digraph $\mathcal{G}_{\sigma(t)}\equiv\mathcal{G}^w$
with all its directed cycles being consistent.  Since any
strongly connected graph contains a directed spanning tree, $\mathcal{G}^w$
contains a directed spanning tree
$\mathcal{T}^d=\bigcup^\varrho_{k=1}\mathcal{T}^d_k$, where the
initial and the terminal node of the directed path $\mathcal{T}^d_k$
are $i_k$ and 1, respectively.  Moreover, the nodes in the directed path from
$i_k$ to 1 are in the order $i_k, i_k-1,\ldots, i_{k-1}+1, 1,$ $1\leq
k\leq \varrho$, $i_0=1$. Associated with the $n$ nodes, we can similarly define 
\begin{align}
q_{1}=1,\;q_{j}=w^{-1}_{(i_{k-1}+1)1}
\prod^{j-1}_{r=i_{k-1}+1}w^{-1}_{(r+1)r}\;
\mbox{for}\;i_{k-1}+1\leq j\leq i_k.\nonumber
\end{align}
Let
$Q={\rm diag}(q_{1},...,q_{n})$ denote the diagonal matrix with diagonal entries $q_1,...,q_n$. It is easy to see that $QL_{\sigma(t)}Q^{-1}\equiv QL_{\sigma(0)}Q^{-1}$
is the Laplacian of the fixed digraph $\mathcal{G}^w$ and
\begin{equation}
\label{aqq} \alpha^TQx(t)
\end{equation}
is time-invariant, where $\alpha=(\alpha_1,...,\alpha_n)^T$ with
$\alpha_i>0,\sum^n_{i=1}\alpha_i=1$ is the left eigenvector of
$QLQ^{-1}$ associated with eigenvalue 0, that is,
$\alpha^TQLQ^{-1}=0$. Similar to the undirected graph case, we
can show the following result, whose proof is omitted due to space limitations.

\begin{them}
\label{them6} In the fixed strongly-connected digraph case, $d^*>0$ if the
initial condition $x(0)$ satisfies
$\big|\alpha^TQx(0)\big|>\sup_{z\in X}|z|$.
\end{them}

\begin{rem}
Clearly, by the relation (\ref{13}) we always have $d^*\leq \max_{1\leq i\leq
n}|x_i(0)|_{X}$. Generally, the final distance $d^*$ between
agents and $X$ depends on the initial conditions, graph
$\mathcal{G}_{\sigma}$, matrix $W$ and the shape of $X$. The
computation of $d^*$ is very complicated and it is not easy to give
its value, or even a lower bound because our connectivity condition
and convex set are quite general.  On the other hand, in some
special cases, we can certainly discuss $d^*$. For example, when
$\mathcal{G}_{\sigma}$ is  undirected, $UJSC$ and $X$ is a ball
with center $(0,0)$ and radius $r_0$, if
$|\textbf{1}^T\widetilde{x}(0)/n|> r_0$ (the sufficient condition in
Theorem \ref{them5} is satisfied),
then
$$d^*\geq\sqrt{2(|\textbf{1}^T\widetilde{x}(0)/n|-r_0)}>0,
$$
because
$\lim_{t\rightarrow\infty}(|x_i(t)|_{X}-|x_j(t)|_{X})=0$ with
$|x_i(t)|_{X}=|\widetilde{x}_i(t)|_{X}$ and
$\lim_{t\rightarrow\infty}|\widetilde{x}_i(t)|_{X}\geq |\textbf{1}^T\widetilde{x}(t)/n|_{X}=|\textbf{1}^T\widetilde{x}(t)/n|-r_0$
with $\textbf{1}^T\widetilde{x}(t)/n\equiv\textbf{1}^T\widetilde{x}(0)/n$.
Similar estimation can also be given for the fixed strongly-connected digraph case.
\end{rem}

\subsection{Special Case: $X=\{(0,0)\}$}

Here we consider a special case when the set becomes a point. Without loss of generality, take
$X=\{(0,0)\}$, which can be regarded as a stationary leader of the multi-agent system. Then system
(\ref{1}) with control law (\ref{2}) can be rewritten as
\begin{equation}
\label{3}
\dot x_i(t)=\sum_{j\in \mathcal{N}_i(\sigma(t))}(w_{ij}x_j(t)-x_i(t))
\end{equation}
or in the compact form: $\dot x(t)=-L_{\sigma(t)}x(t)$, where
$L_{\sigma}$ is given in (\ref{lap}).

\begin{rem}
System (\ref{3}) is a generalized model for various models in the
multi-agent literature. For example, when $w_{ij}=1$ for $(i,j)\in
\mathcal{E}^w$, system (\ref{3}) becomes the standard consensus model
with all connection weights equal to 1. Moreover, the bipartite
consensus model discussed in \cite{alt} is a special case of system
(\ref{3}) with $w_{ij}=1$ or $-1$.
\end{rem}

\begin{rem}
Different from the feedback control
$u_i=\sum_{j\in\mathcal{N}_i}w_{ij}(x_j- x_i)$ given in \cite{lin1,lin2}, our distributed
control is $u_i=\sum_{j\in \mathcal{N}_i}(w_{ij}x_j-x_i)$. As stated
in \cite{lin1}, the system matrix generated by $v_i$ may have
eigenvalues with positive real parts and then the resulting system may be
unstable. Here if the graph $\mathcal{G}_{\sigma}$ is undirected and switching
(or fixed and strongly connected) with consistent directed cycles
of $\mathcal{G}^w$, then all the eigenvalues of
$L_{\sigma}$ (or $L$) have non-negative real parts, which implies that
for the two cases system (\ref{3}) is always stable.
\end{rem}

Consider system (\ref{3}) associated with a \emph{UJSC} undirected
graph $\mathcal{G}_{\sigma}$. Recalling the diagonal
matrix $P$ in (\ref{pp}), we first have the following theorem.

\begin{them}\label{them2}
For system (\ref{3}), if the undirected communication graph
$\mathcal{G}_{\sigma}$ is $UJSC$ and all directed cycles of
the configuration graph $\mathcal{G}^w$ are consistent, then, for any initial condition
$x_i(0),i=1,...,n$,
\begin{align}
\lim_{t\rightarrow\infty}x_i(t)=\frac{\sum^n_{j=1}p_{j}x_j(0)}{np_{i}},\quad
1\leq i\leq n.\nonumber
\end{align}
\end{them}

\emph{Proof:} Recalling the notations $\widetilde{x}(t)$ and
$\widetilde{L}_{\sigma(t)}$ used in Subsection III. B, we have $\dot{
\widetilde{x}}(t)=-\widetilde{L}_{\sigma(t)}\widetilde{x}(t)$. According to
Theorem 2.33 in \cite{renbook}, for any $x_i(0),\;1\leq i\leq n$,
$$
\lim_{t\rightarrow\infty}\widetilde{x}_i(t)=\frac{\sum^n_{j=1}
\widetilde{x}_j(0)}{n},\;
1\leq i\leq n,
$$
which implies the conclusion.
\hfill$\square$

Next we consider system (\ref{3}) with a fixed strongly connected digraph
$\mathcal{G}_{\sigma(t)}\equiv\mathcal{G}^w$.
Clearly, $L$ is diagonally
dominant and all its eigenvalues are either 0 or with positive real
parts.

\begin{lem}
\label{lem4} 0 is an eigenvalue of $L$ if and only if all directed cycles of $\mathcal{G}^w$ are consistent.
\end{lem}

\emph{Proof:} Sufficiency.
If all directed cycles of $\mathcal{G}^w$
are consistent, then by the discussions in Subsection III. B, there is an invertible diagonal matrix $Q$ such that
$QLQ^{-1}$ is the Laplacian of digraph $\mathcal{G}^w$. Since all row sums of any Laplacian are zero,
any Laplacian has an eigenvalue zero. The sufficiency follows from that similar matrices have the
same eigenvalues.

Necessity. Let us show it by contradiction. Suppose that
$\mathcal{G}^w$ contains inconsistent directed cycles. On one hand, by
Theorem \ref{them1} (ii),
$\lim_{t\to\infty}x_i(t)=0$, $i=1,...,n$ for any initial conditions (noticing that $X=\{(0,0)\}$).
On the other hand, let $\xi\neq 0$ be the eigenvector of $L$ with eigenvalue 0, that is, $L\xi=0$. Clearly,
$x(t)\equiv\xi$ for initial condition $x(0)=\xi$,
which yields a contradiction. Thus, the necessity follows.
\hfill$\square$

Recalling the matrix $Q$ and vector $\alpha$ defined in (\ref{aqq})
along with Lemma \ref{lem4} and Theorem 2.13 in \cite{renbook}, we have the following theorem.

\begin{them}
\label{them4} Consider system (\ref{3}) with a fixed strongly
connected digraph $\mathcal{G}_{\sigma(t)}\equiv\mathcal{G}^w$. Then $\lim_{t\rightarrow\infty}
x_i(t)$ $=0$, $i=1,...,n$ for any initial conditions if and only if
$\mathcal{G}^w$ contains inconsistent directed cycles. Moreover, if all directed
cycles of $\mathcal{G}^w$ are consistent, then, for any initial condition
$x_i(0),i=1,...,n$,
$$
\lim_{t\rightarrow\infty}x_i(t)=\frac{\sum^n_{j=1}\alpha_jq_{j}x_j(0)}{q_{i}},\; 1\leq i\leq n.
$$
\end{them}

\begin{rem}
Theorem \ref{them1} stated that, in the fixed graph case, if there
exist irregular weak cycles in $\mathcal{G}$, then, for any initial
condition, all agents converge to the target set (which is the
origin here), which implies that all eigenvalues of generalized Laplacian
$L$ have positive real parts. In fact, as shown in Theorems
\ref{them2} and \ref{them4}, the converse is also true. Theorems
\ref{them2} and \ref{them4} provided necessary and sufficient
conditions to characterize the relationship between the stability of
system (\ref{3}) and the properties of graph $\mathcal{G}$.
\end{rem}

\begin{rem}
Clearly, the results in Theorems \ref{them2} and \ref{them4} are
consistent with the conventional results in \cite{olf,renbook,alt}.
In fact, if all $w_{ij}$'s are $1$, both $P$ and $Q$ given in Subsection III. B
are the identity matrix, which implies that all agents will achieve
a consensus for any initial conditions by the conclusions in
Theorems \ref{them2} and \ref{them4}. Moreover, Theorem 2 in
\cite{alt} showed that all agents will converge to the origin for the
structurally unbalanced graph case or achieve the bipartite
consensus for the structural balanced graph case, which can be
obtained from Theorem \ref{them4} in this paper by noticing that a
digraph with all configuration weights being $-1$ or $1$ is structurally balanced if
and only if all its directed cycles are consistent. Due to the convex set
and complex-value weights, the method given in \cite{alt} for the
bipartite consensus cannot be applied
directly to solve our problem.
\end{rem}

Sometimes, we need to check whether the weak cycles in the configuration
graph are consistent and it is known that the consistency of weak
cycles in digraphs implies that of directed cycles. In the strongly
connected digraph case, the converse is also true and then we only
need to check the consistency of all directed cycles instead of that
of all weak cycles as the next result shows.

\begin{them}
\label{lem5}
Suppose $\mathcal{G}^w$ is strongly connected. Then all directed cycles of $\mathcal{G}^w$ are consistent if and only if all weak cycles of $\mathcal{G}^w$ are consistent.
\end{them}

\emph{Proof:} The sufficiency part is straightforward. We focus on the necessity part.
Without loss of generality, let the weak cycle in $\mathcal{G}^w$ take
the following form
$$i_1e_1i_2e_2\cdots e_{k_1-1}i_{k_1}e^{-1}_{k_1}i_{k_1+1}\cdots i_ke^{-1}_ki_1,$$
 where $e_{r}=(i_r,i_{r+1}),$ $r=1,...,k_1-1$; $e^{-1}_{r}=(i_{r+1},i_r),$ $r=k_1,...,k, i_{k+1}=i_1$.
Since $\mathcal{G}^w$ is strongly connected, for each $r=k_1,...,k$, there is a directed path
$\mathcal{P}_r$ from $i_r$ to $i_{r+1}$. Because the
directed cycles $\mathcal{P}_re^{-1}_r,r=k_1,...,k$ are
consistent, $w(\mathcal{P}_r)w_{i_{r+1}i_r}=1$, where
$w(\mathcal{P}_r)$ is the product of all configuration weights on directed path
$\mathcal{P}_r$. Then, from the consistency of the directed cycle
$i_1e_1i_2e_2\cdots e_{k_1-1}\mathcal{P}_{k_1}\cdots \mathcal{P}_k$, we have
\begin{align}
w_{i_1i_2}&\cdots w_{i_{k_1-1}i_{k_1}}w^{-1}_{i_{k_1+1}i_{k_1}}\cdots w^{-1}_{i_{k+1}i_{k}}\nonumber\\
&=w_{i_1i_2}\cdots w_{i_{k_1-1}i_{k_1}}w(\mathcal{P}_{k_1})\cdots w(\mathcal{P}_{k})=1.\nonumber
\end{align}
Thus, the conclusion follows. \hfill$\square$

\section{Numerical Example}

In this section, we provide an example to illustrate the results obtained in this paper.

Consider a network of five agents with node set
$\mathcal{V}=\{1,2,3,4,5\}$ and the complex-value adjacency matrix
$W=(w_{ij})$. The convex set to be surrounded is the unit ball in
$\mathbb{R}^2$. The initial conditions are $x_1(0)=2+4\iota$,
$x_2(0)=4+3\iota$, $x_3(0)=-4-3\iota$, $x_4(0)=-4+2\iota$,
$x_5(0)=2+3\iota$ (marked as $\circ$ in Figures 1 and 2).

Two cases are shown as follows:
\begin{itemize}
\item Consistent case: Take $w_{12}=w_{23}=w_{34}=e^{\frac{\pi}{2}\iota}$,
$w_{45}=e^{\frac{\pi}{3}\iota}$, $w_{51}=e^{\frac{\pi}{6}\iota}$, and
all other configuration weights are zero. Then the resulting configuration graph is $\mathcal{G}^w=(\mathcal{V},\mathcal{E}^w)$
with arc set $\mathcal{E}^w=\{(1,2),(2,3),(3,4),(4,5),(5,1)\}$.
The communication graph of the multi-agent system
is periodically switched between two graphs
$\mathcal{G}_1=(\mathcal{V},\mathcal{E}_1),\mathcal{G}_2=(\mathcal{V},\mathcal{E}_2)$
with $\mathcal{E}_1=\{(1,2),(3,4),(5,1)\}$ and
$\mathcal{E}_2=\{(2,3),(4,5)\}$ in the following order:
$\mathcal{G}_1,\mathcal{G}_2,\mathcal{G}_1,\mathcal{G}_2,\ldots$
with switching period $5$. Clearly, $\mathcal{G}_{\sigma}$ is $UJSC$
and all directed cycles of $\mathcal{G}^w$ are consistent. Figure 1
demonstrates that all agents accomplish the consistent set surrounding
at time $t=2000$, where the five agent trajectories are described by
the solid lines and the projection vectors of the final positions of
the agents are described by dashed lines.

\item Inconsistent case: Take $w_{12}=w_{23}=w_{34}=e^{\frac{\pi}{2}\iota}$,
$w_{45}=e^{\frac{\pi}{3}\iota}$, $w_{51}=e^{\frac{\pi}{3}\iota}$,
$w_{14}=e^{-\frac{\pi}{2}\iota}$, and all other configuration weights are zero.
Suppose the communication graph is fixed, that is,
$\mathcal{G}_{\sigma}\equiv\mathcal{G}^w=(\mathcal{V},\mathcal{E}^w)$ with
$\mathcal{E}^w=\{(1,2),(2,3),(3,4),(4,5),(5,1),(1,4)\}$.  Note that the configuration graph $\mathcal{G}^w$
defined by the new configuration weights is clearly inconsistent.
Figure 2 shows that all agents converge to the unit ball, where the final positions of the
five agents at time $t=2000$ are marked with $*$.
\end{itemize}

\begin{figure}[!htbp]
\centering
\includegraphics[width=3.6in]{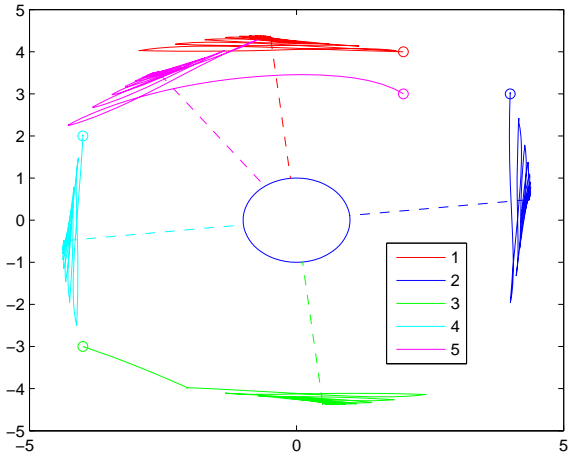}
\caption{The consistent cycles yield the consistent set surrounding.}
\end{figure}

\begin{figure}[!htbp]
\centering
\includegraphics[width=3.6in]{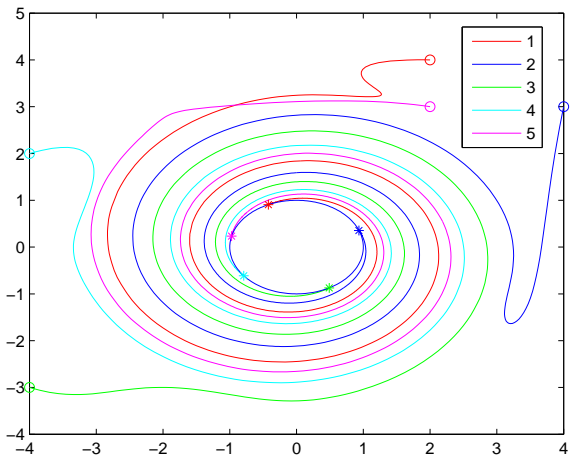}
\caption{The inconsistent cycles yield the inconsistent set surrounding.}
\end{figure}

\section{Conclusion}
In this paper, we proposed a formulation and a distributed controller for
set surrounding problems.  We discussed both consistent and inconsistent cases, and
obtained the necessary/sufficient conditions for multi-agent systems
with communication topologies described by joint-connected graphs.
Moreover, we showed when the consistent case
can be guaranteed, and also provided conditions on
the leader-following consensus when the target
set becomes one point.

 Many interesting problems about set surrounding of multi-agent systems remain to be done, such as the analysis in the case with various uncertainties, the construction of configuration graphs based on given optimization indexes, and surrounding control design for agents moving in a high-dimensional space with the help of rotation matrices.


%

%

%
%

\ifCLASSOPTIONcaptionsoff
  \newpage
\fi



%

\section*{Acknowledgment}

The authors would like to thank Dr. Anton V. Proskurnikov and Prof. Ming Cao
for their comments on the Babalat's Lemma for switching cases.

%

%






\end{document}